# Niobium-based sputtered thin films for Corrosion Protection of proton-irradiated liquid water targets for [$^{18}$F] production

H. Skliarova[1,2], O. Azzolini[1], O. Dousset[1,2], R.R. Johnson[3], V. Palmieri[1,2]

[1] INFN, Laboratori Nazionali di Legnaro, Legnaro (Padova), Italy.
[2] Università di Padova, Padova, Italy.
[3] BEST Cyclotron Systems Inc., Vancouver BC, Canada.

## Abstract

Chemically inert Coatings on Havar® entrance foils of the targets for [$^{18}$F] production via proton irradiation of enriched water at pressurized conditions are needed to decrease the amount of ionic contaminants released from Havar®. In order to find the most effective protective coatings, the Nb-based coating microstructure and barrier properties have been correlated with deposition parameters as: substrate temperature, applied bias, deposition rate and sputtering gas pressure. Aluminated quartz used as a substrate allowed us to verify the protection efficiency of the desirable coatings as diffusion barriers. Two modeling corrosion tests based on the extreme susceptibility of aluminum to liquid gallium and acid corrosion were applied. Pure Niobium coatings have been found less effective barriers than Niobium-titanium coatings. But Niobium oxide films, according to the corrosion tests performed, showed superior barrier properties. Therefore Multi-layered Niobium-Niobium oxide films have been suggested, since they combine the high thermal conductivity of Niobium with the good barrier properties of Niobium oxide.

## Introduction

Positron emission tomography (PET) is a nuclear medical imaging technique that produces a three-dimensional image or map of functional processes in the body. To conduct the scan a short-lived positron-emitting radionuclide is introduced into the body in a form of labeled bioactive molecule. Nowadays the most commonly used radiotracer in clinical PET scanning is [$^{18}$F]fluoro-2-deoxy-D-glucose ([$^{18}$F]FDG), an analogue of glucose that is labeled with $^{18}$F. This radiotracer is used in essentially all scans for oncology and most scans in neurology, and thus makes up the large majority of all of the radiotracers (>95%) used in PET and PET/CT scanning.

There are several routes for the production of reactive [$^{18}$F][1], however, the greatest production yield of non-carrier-added high specific activity [$^{18}$F]F$^-$ is attained by proton irradiation of enriched water targets via $^{18}$O(p,n)$^{18}$F reaction[2]. Since the process is realized by cyclotron accelerated proton beam interaction with small volume [$^{18}$O]H$_2$O targets at pressurized conditions there are strict demands for suitable materials for both the target chamber and the entrance foil of a high-power target[3].

Taking into account that the accelerated proton beam generates high temperatures in the target chamber, the target materials must have high melting point and high thermal conductivity to guarantee the stability and efficient heat dissipation. As soon as the routine production of [$^{18}$F]F$^-$ is carried out at overpressure of 4 MPa (580 psi)[4], the target materials require a high mechanical strength. The entrance foil appears to be the "Achilles' heel" of the high power liquid target: it must be as thin as possible to not absorb the proton beam energy, but it must resist to high pressure.

It is well known that the aqueous corrosion of chamber wall can be accelerated both by the direct interaction of the beam with the wall materials and by the exposure to the media radiolysis products.

The water radiolysis models for accelerators[5] predict next species to be formed in environments:

$$H_2O \rightarrow OH, H, H_2, H_2O^+, H_2O_2, e^-_{aq}, ....$$

With the exception of hydrogen peroxide (H$_2$O$_2$), oxygen (O$_2$), and hydrogen

(H), the lifetime of the majority of radiolysis products, is short, on the order of microseconds, and the steady state concentration is typically on the order of $10^{-10}$ to $10^{-8}$ M. Therefore, these short lived species play an important role in the corrosion mechanism at a surface directly impinged by the high energy proton beam, where the steady state concentration will be high while the beam is on. The formation of ionic contaminants acquired via metal corrosion by proton irradiated water has been found to affect both the [$^{18}$F]F$^-$ reactivity and the labeling yield of radiopharmaceuticals [6,7,8]. So the chemical inertness of materials to select for the target body and the entrance foil fabrication is very important.

High tensile strength (1860 MPa), high melting point (1480 °C), and moderate thermal conductivity (14.7 Wm$^{-1}$K$^{-1}$ at 23 °C) of Havar®, a non-magnetic alloy (Co 42%, Cr 19.5%, Fe 19.3 %, Ni 2.5%, W 2.6%, Mo 2.2%, Mn 1.7%, and C 0.2%), have brought it great popularity as an entrance foil material. The main disadvantage of Havar® as the entrance foil is a wide variety of radionuclidic and chemical impurities[8] generated while exposed to proton irradiated water. Refractory metals, such as Niobium (Nb), platinum (Pt), zirconium (Zr) and tantalum (Ta) have high melting point and excellent chemical resistance. Though their low yield point in the stress strain curve do not allow to use them for fabrication of the entrance foil for routine production of [$^{18}$F] under high pressure conditions[9]. A combination of high tensile strength Havar® substrate top-coated by protective corrosion-resistant barriers has been proven to be a solution. Actually it was found that the use of sputtered Nb coating on Havar® entrance foils decreases the amount of ionic long-lived impurities more than ten times[3].

The complex shape of the entrance foil optimized to match the exact curvature of the beam has been found to reduce the power density in the beam strike area. Therefore the use of dome-shaped entrance foils (Figure 1) facilitates the dissipation of the heat generated when the beam passes through the entrance foil. For that reason, high uniformity of the coating thickness on such a complex shape of the substrate is mandatory.

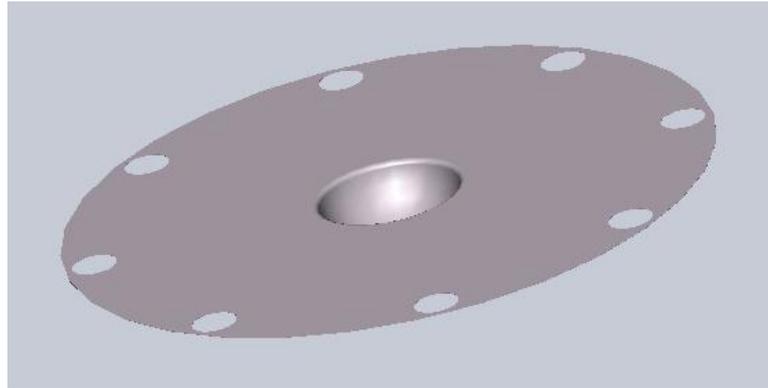

Figure 1 Dome-shaped entrance foil for cyclotron target

Defects, such as pores, pin-holes and grain boundaries are a primary limitation for thin film barriers. Even if the bulk material of a chosen barrier film met all the requirements of a diffusion barrier, the presence of the grain boundaries and other extended structural defects in the thin film could entirely annihilate its usefulness as a barrier[10].

1. **Literature approaches for Corrosion resistance films**

In a case of thin films the atomic diffusion is no longer determined by regular bulk processes. For a thin film, indeed, grain boundaries and other micro-structural defects assume a more important role. Grain boundaries and extended defects offer paths which can carry foreign atoms rapidly across the films thereby dominating the atomic traffic[10]. On the other hand, diffusion through the amorphous layers is very difficult due to the irregularity of the atomic structure[11]. The high corrosion resistance of amorphous alloys is explained by the formation of uniform passive alloys without weak points. The reason is that such passive layers are free of the typical structural defects of a crystalline state, such as dislocations, grain boundaries, second phases, precipitates and segregates[12]. Thereby sputtered Niobium-based amorphous alloy films with cromium[12,13,14], molybdenum[12,15] and tungsten[16] have been found more corrosion resistant to concentrated hydrochloric acid than single element crystalline films.

Several approaches in order to obtain amorphous metallic film have been also applied in the past, such as deposition at low temperature, alloying with other elements, gas admixture, deposition at high deposition rate and use of a substrates with amorphous structure.

Cooling of a substrate at cryogenic temperatures during sputter deposition decreases the mobility of atoms arriving to the substrate. Whereas the heat of condensation is absorbed without essential rise in temperature, crystalline growth is inhibited resulting in amorphous structure. Thus aluminum and aluminum-copper alloy amorphous like films have been deposited at temperatures lower 30 K using He, Ne, Ar as sputtering gases[17].

As far as the monoatomic amorphous films are not stable at room temperatures, alloying is used to increase the thermodynamical stability of a system. Thus sputtered films of Niobium and manganese alloys with 40÷75 Nb content have been found amorphous by XRD investigation, whereas pure Nb and pure Mn films were crystalline[18].

Admixture of different gases, such as helium[19], hydrogen[20] and oxygen[21] in a metal has been found to decrease the degree of film crystallinity. Thus amorphous Ni, Fe, Au films were deposited by getter-sputtering using both approaches: low deposition temperature, and the incorporation of large amounts of helium gas into the film[19]. Amorphous Yb-H films with 40-55% $H_2$ contamination were deposited by vapor condensation in extra high vacuum[20]. Amorphous-like tungsten films were prepared by oxygen admixture during sputtering[21].

Increase of the deposition rate, maintaining cooling of a substrate, can reduce the adatom migration mobility, resulting in a less ordered structure than the one obtained by conventional deposition methods[22].

The morphology of a substrate surface also can be used to promote film growing in desirable way. Thus Ag, Au, Cu amorphous coatings were fabricated by thermal vacuum deposition onto a dielectric polymeric substrate at room temperature[23].

## 2. Experimental techniques and apparatus

The deposition was carried out by 2 inches in diameter planar cathode magnetron sputtering unbalanced source of the II type in the Window and Savvides classificatione[24]. Depositions have been done in a cylindrical, 316 L stainless steel vacuum chamber of 11 cm diameter and 26 cm in length. Base pressure of $5 \cdot 10^{-6}$ mbar could be reached by a Pfeiffer turbo molecular pump of 360 l/min and Varian Tri Scroll Pump 12 m$^3$/hr as a primary pump.

In order to provide the coating uniformity onto dome-shaped substrates, deposition at different Target-Substrate distances (6 cm and 12 cm), has been carried out. SEM pictures of coated sample cross sections displayed the thickness distribution of the film. The 6 cm distance from the Cathode to the Substrate has been chosen as the most suitable for a uniform coating thickness. Thus the routine depositions were carried out on different planar substrate holders with possibility to apply bias, to cool down or to heat up the substrate, keeping constant the distance from the cathode at 6 cm.

High grade bulk Nb for superconducting cavities (99.99% purity, 250 Residual Resistivity Ratio) was used for target fabrication. Argon (99.9999% purity) was used as process gas for Nb thin films deposition. Niobium oxide films were obtained by reactive magnetron sputtering in the same sputtering system using a mixture of argon and oxygen (99.9999% purity) gases.

As far as the direct test of a protective coating on Havar® entrance foil on a cyclotron facility is not a routine operation and it can be performed only on final samples. So it was better to find at first the best recipes for the corrosion protection barriers by more simple preliminary corrosion tests. In order to verify the efficiency of obtained coatings as diffusion barriers, aluminum was chosen as a substrate. Being extremely susceptible for acid corrosion and liquid gallium corrosion, aluminum was considered to be a good substrate for our corrosion modeling.

We used aluminum substrate samples for two tests of the barrier quality of our coatings: 1) the acid solution test; 2) the test with liquid gallium.

$H^+$ is the smallest particle that could be obtained chemically, and it will penetrate easily through whatever prolonged defects in a protective film. Thus both acid solution tests and Liquid gallium tests can be used for discarding those sputtering recipes that gave rise to poor barrier coatings. We think that if a film does not overcome our corrosion tests, will most probably perform badly under proton irradiation. Then we make a further assumption that if the film could protect aluminum from both liquid Ga penetration and Hydrochloric acid, there will be higher probability that it will work as a barrier for proton irradiated water also.

The use of aluminum foil substrates has met some problems: a first limit came from the high flexibility of the foil; then the back-penetration into the substrate of silver from the silver-paint used to bond the sputtering substrate to the sample holder. Silver paint is a very efficient bonder, but it is difficult to remove from the sample holder without damage. So we chose the sputter deposition of aluminum onto quartz as our main technique of producing the substrates for Nb-based coatings.

Aluminum films of about 2.5 μm thickness were deposited in serial onto 9×9×1 mm optical finishing fused quartz substrates by DC-magnetron sputtering of Al target at $3.3·10^{-3}$ mbar Ar pressure and $I_{DC}$=0.25 A using a grounded substrate holder with 6 cm distance to the cathode.

Nb coatings were also sputtered onto the diamond-machined mirror-like aluminum bulk substrates for comparison. The tests had been carried out on both the aluminated quartz and on the mirror-like bulk aluminum. Coatings on both substrates give equivalent results.

Quartz samples 9×9×1 mm have been also sputtered for XRD investigation simultaneously with the aluminated quartz samples for corrosion tests. The XRD investigation was carried out by performing the $\theta$-$\theta$ Gonio scan in Bragg-Brentano configuration with $2\theta$ from 10° to 100° (180°-$2\theta$ is the angle between the X-ray source, the substrate and the detector). The 1.54 Å Cu-Kα X-ray was used to observe X-ray diffraction of thin films with a Panalytical (ex-Philips) PW3040/60 Diffractometer. The data of the diffracted beam intensity dependence on $2\theta$ have been plotted and fitted by the X'Pert Highscore software in order to obtain the peak position and the integral breadth.

The shape of the graphs has been used to provide a general recognition of crystalline or amorphous structure. Thus the graphs that display the exact sharp peaks indicate a crystalline structure, while the peak absence signifies that the film presents an amorphous structure. The average grain size has been determined from the plot of $I_r$ versus $2\theta$ (where $I_r$ is the intensity of the diffracted X-ray and 180°-$2\theta$ is the angle between the X-ray source, the substrate and the detector) using Debye-Scherrer formula:

$$\qquad$$

Being λ the wave length of the X-ray source, in case of Cu-Kα X-ray λ=1.54056Å and $B$ the Bragg XRD peak breadth. The cubic lattice parameter has been obtained due to the formula:

$$\qquad$$

Where $h, k, l$ are the Miller indexes; and $d_{hkl}$ is the spacing between the planes in the atomic lattice evaluated by use of Bragg law:

$$\qquad$$

The acid test has been carried out by immersing the samples into a solution of 10% hydrochloric acid at the temperature 30±5° C during 10 min. The results have been evaluated due to the number of hydrogen bubbles appeared on the surface of the coating. The acid test results have been evaluated in a scale from "1" to "5". The value "1" (Figure 1, a) means the smallest amount of hydrogen bubbles and appropriately the best barrier quality of the coating. The value "5" (Figure 2, b) on the contrary means a huge amount of bubbles and the lowest barrier quality.

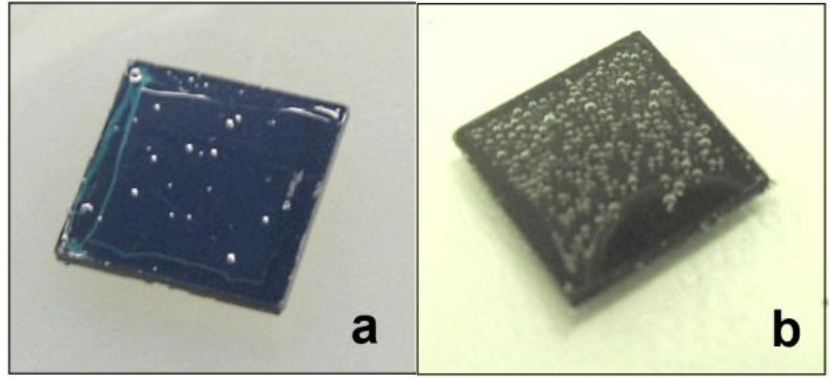

Figure 2 Acid solution resistance: a) high, b) low

The liquid gallium test has been carried out by heating the quartz aluminated samples coated with investigated Nb-based coatings with liquid gallium droplets during 30 hours. The temperature has been kept at 200±10 °C during the whole test. In case of corrosion (Figure 3) liquid gallium has penetrated through the film and the aluminum under-layer has become liquid. In case of corrosion absence (Figure 4), the Ga drop rolls over if the substrate is tilted. In that case it can leave or not Ga traces on the Nb film depending on wettability.

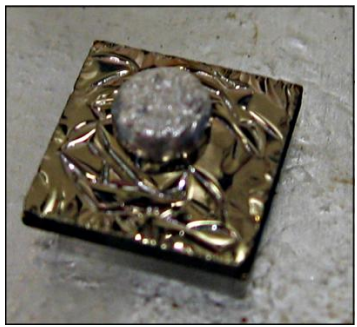 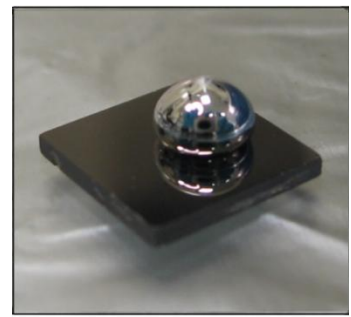

Figure 3 Liquid gallium test: corrosion,   Figure 4 Liquid gallium test: resistance

Wettability is the ability of a liquid to maintain the contact with a solid surface, resulting from intermolecular interactions when the two are brought together, and it is determined by the force balance between the adhesive energy of the surface and the cohesive energy of the liquid. High wettability corresponds to a low contact angle of a droplet and low wettability corresponds to a high contact angle respectively.

## 3. Experimental results (Discussion)

In order to select the best configuration for a uniform deposition onto the dome-shaped entrance foils, Nb film sputtering at $I_{DC}= 0.5$ A using $4.3 \cdot 10^{-3}$ mbar Ar working pressure was carried out onto two stainless steel replica substrates disposed at 6 cm and 12 cm distance from the cathode. The deposition took 1 hour for the 6 cm for the substate-cathode distance and and 2 hours for the 12 cm one. In order to estimate the real thickness of the coating all over the substrate, it was put inside a plastic container, filled with the specific resin for the SEM analysis, mechanically cut, lapped with fine abrasives and chemically etched by Buffered chemichal polishing (BCP), HF : $HNO_3$ : $H_3PO_4$ in ratio 1 : 1 : 2. The replica cross-section was observed by Fei (ex Philips) SEM XL-30 all over the sample, attending to 7 fixed crytical points (Figure 5).

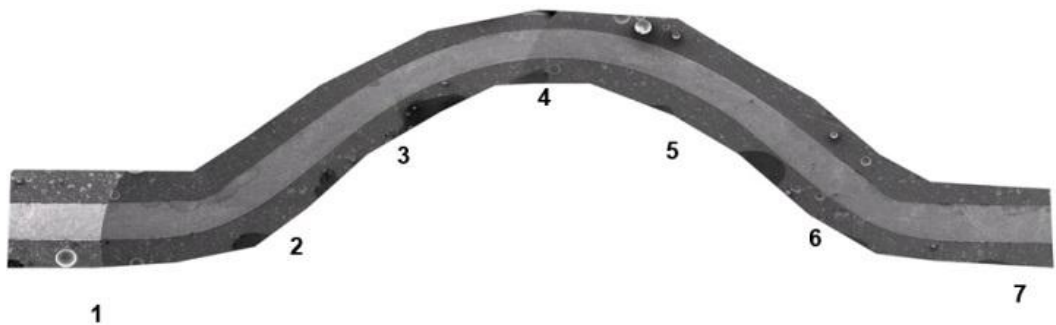

**Figure 5 Critical points of coating thickness onto dome-shaped substrate**

Sputtering of Nb film at 6 cm Substrate-Cathode distance (Figure 7) has been found to be more uniform and also possess more than 2 times higher deposition rate than at 12 cm (Figure 6). So the Substrate-Cathode distance for all subsequent experiments was chosen 6 cm.

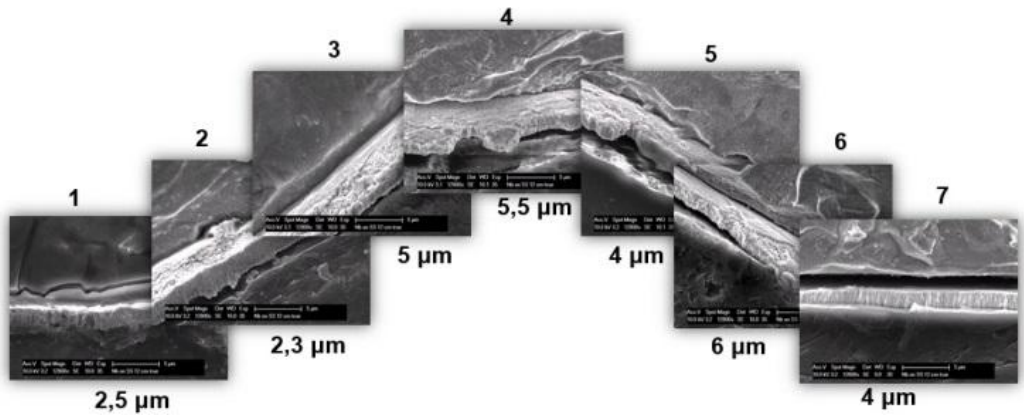

**Figure 6 Thickness distribution of Nb film deposited at 12 cm Substrate-Cathode distance**

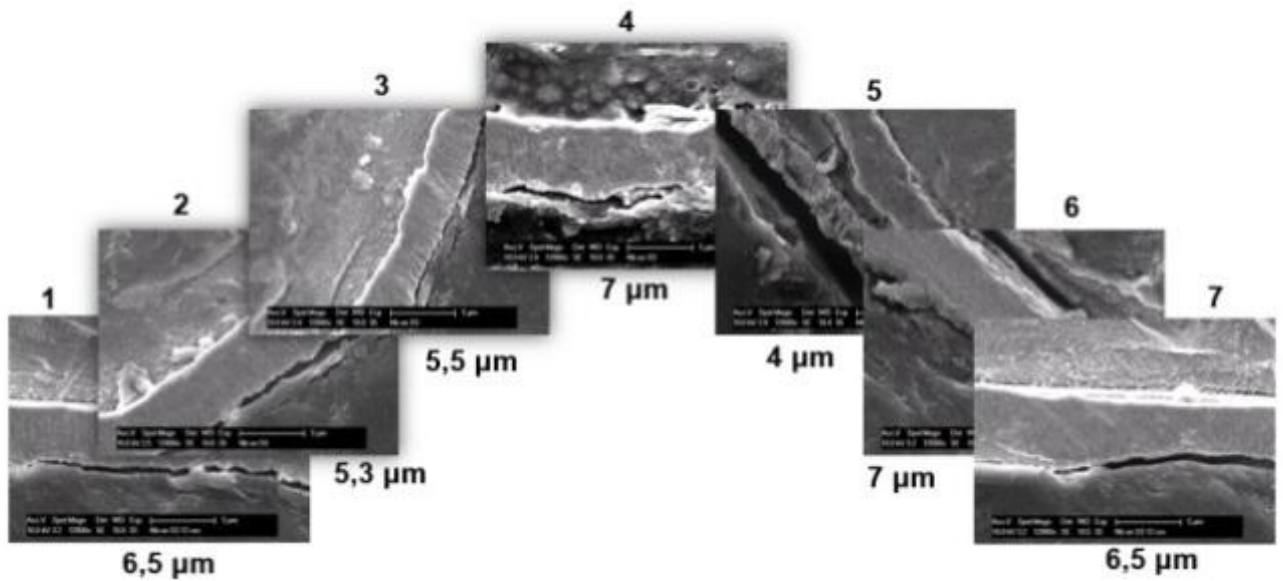

**Figure 7 Thickness distribution of Nb film deposited at 6 cm Substrate-Cathode distance**

## *3.1 Pure Niobium sputtering*

Pushed by the intention to sputter the purest films of Niobium, the most convenient choice appeared to apply the technology from the scientific community of thin film superconducting radiofrequency cavities for particle accelerators [25]. For Niobium Sputtered Copper cavities, high RRR Niobium coatings are prepared by applying to the substrate either a negative bias or a high temperature. Negative bias is well-known to compact the structure, to enlarge the grains and to promote desorption of weakly bounded impurities. High Temperatures promote the grain growth due to the increased adatom mobility.

The influence of a negative bias application to the substrate was tested, without temperature control, from $3·10^{-3}$ mbar at $3·10^{-2}$ mbar (Table 1).

**Table 1 Influence of pressure and negative bias**

| Sample Number | Nb 39 | Nb 29 | Nb 38 | Nb 13 | Nb 7 | Nb 41 | Nb 86 |
|---|---|---|---|---|---|---|---|
| **Bias, V** | grounded | -80 | grounded | -50 | -80 | -80 | -150 |
| **Ar pressure, mbar** | $3·10^{-2}$ | $3·10^{-2}$ | $3·10^{-3}$ | $5·10^{-3}$ | $6·10^{-3}$ | $3·10^{-3}$ | $3·10^{-3}$ |
| $I_{DC}$, A | 0.5 | 0.5 | 0.5 | 0.5 | 0.5 | 0.5 | 0.5 |
| Deposition rate, nm/sec | 0.4 | 0.4 | 0.8 | 0.8 | 0,9 | 0.7 | 0.8 |
| Average grain size, Å | 110 | 120 | 170 | 170 | 170 | 190 | 220 |
| Cubic lattice parameter, Å | 3.334 | 3.316 | 3.325 | 3.312 | 3.312 | 3.318 | 3.319 |
| Liquid Ga corrosion | corroded | corroded | resists | resists | resists | resists | resists |
| Liquid Ga wettability | low | high | low | high | high | high | high |
| Acid test | 5 | 3 | 5 | 2 | 1 | 1 | 2 |

From table 1 we see that at low Argon pressure, the effect of Bias increases the grain size. Increasing the Argon pressure from the range $10^{-3}$ mbar to the range $10^{-2}$ mbar, the ion bombardment of the growing film is decreased, due to more collisions with the argon gas. Indeed the application of negative bias of -80 V to the substrate holder at $3·10^{-2}$ mbar of argon (*sample Nb 29*) results in a negligible effect on the film micro-structure if compared to the grounded one (*sample Nb39*). Even the average grain size measured by the X-ray Diffractometer does change passing from a grounded substrate to a negatively biased one.

The cubic lattice parameter of negative biased coatings is lower than the grounded ones, and more to 3.300 Å, value for bulk Niobium (Figure 8). Nevertheless the cubic lattice parameter value obtained for -80V biased coating (*Nb 29*) is lower than grounded, and this is a sign of lattice compression however due to the ion bombardment.

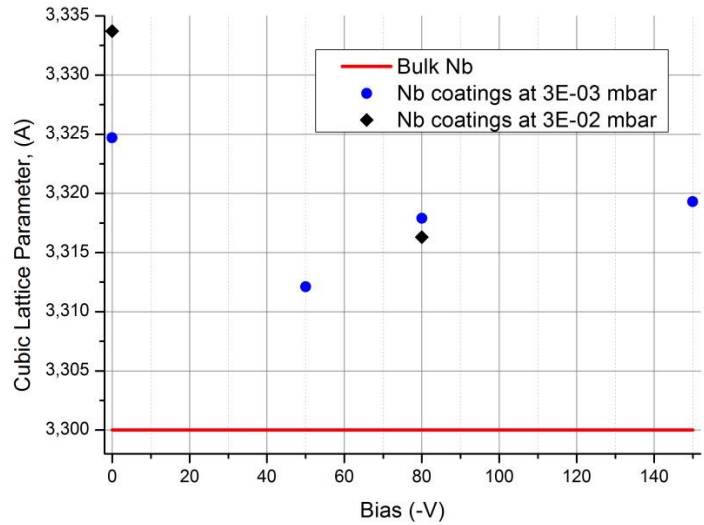

Figure 8. Negative bias influence on Nb film cubic lattice parameter

Nevertheless, it must be point out that, while for a diode sputtering, a negative bias is always effective, for a magnetron sputtering process the effectiveness of negative bias depends on the degree of immersion of the substrate into the plasma, in other words depends on the type of magnetic confinement, cathode-substrate distance and argon pressure.

According to liquid gallium and acid tests, the best condition for pure Nb corrosion protection coating are the following: -80 V bias and $3·10^{-3}$ mbar argon pressure.

Negative bias increases the liquid Ga surface wettability of the coating. In the meantime, Negative bias promotes grain growth. Large grains mean large grain boundaries. Hence one would be led to put in connection liquid metal wettability with large grains or large grain boundaries.

In order to test the substrate temperature effect, Nb films were deposited at high temperature (300 °C, 400 °C, 500 °C) (Table 2). According the know-how of superconducting cavities, high temperatures increase the purity of Niobium films and provide larger grain size and a crystal lattice parameter close to the standard bulk value of 3,300 Å (Figure 9).

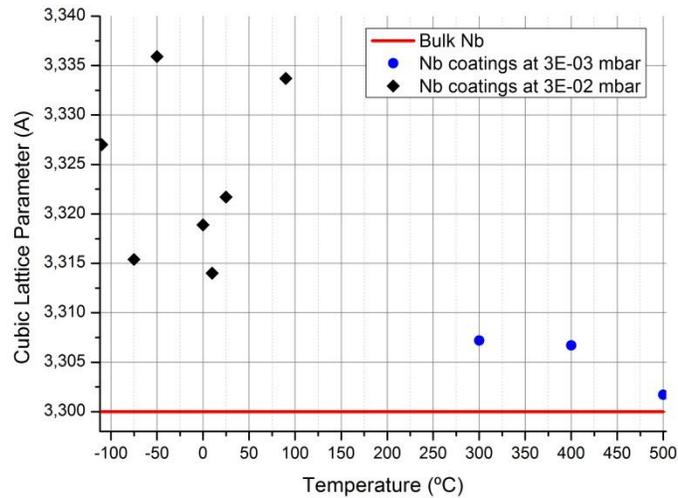

Figure 9 Temperature influence on Nb film cubic lattice parameter

If we call $T_m$ a material melting point, the parameter that rules all the main metallurgical properties is the homologous temperature[26], $\theta_h = T_{depos}[K]/T_m[K]$. Since the Nb melting temperature is 2468°C, a temperature as 500 °C is not enough for providing the bulk-like microstructure.

Bigger grains mean also bigger grain boundaries, but those are not acceptable as corrosion barriers.

Table 2 Influence of temperature

| Sample Number | Nb 39 | Nb 19 | Nb 14 | Nb 16 | Nb 27 | Nb 35 | Nb 36 | Nb 34 | Nb 32 | Nb 33 |
|---|---|---|---|---|---|---|---|---|---|---|
| **Temperature, °C** | floating | 500 | 400 | 300 | 25 | 10 | 0 | -50 | -75 | -110 |
| Ar pressure, mbar | $3 \cdot 10^{-2}$ | $7 \cdot 10^{-3}$ | $7 \cdot 10^{-3}$ | $4 \cdot 10^{-3}$ | $3 \cdot 10^{-2}$ | $3 \cdot 10^{-2}$ | $3 \cdot 10^{-2}$ | $3 \cdot 10^{-2}$ | $3 \cdot 10^{-2}$ | $3 \cdot 10^{-2}$ |

| $I_{DC}$, A | 0.5 | 0.5 | 0.5 | 0.5 | 0.5 | 0.5 | 0.25 | 0.5 | 0.5 | 0.5 |
|---|---|---|---|---|---|---|---|---|---|---|
| Deposition rate, nm/sec | 0.4 | 0.8 | 0.7 | 0.7 | 0.7 | 0.6 | 0.3 | 0.7 | 0.6 | 0.5 |
| Average grain size, Å | 110 | 230 | 200 | 180 | 80 | 70 | 60 | 80 | 70 | 70 |
| Cubic lattice parameter, Å | 3.334 | 3.302 | 3.307 | 3.307 | 3.322 | 3.314 | 3.319 | 3.336 | 3.315 | 3.327 |
| Liquid Ga corrosion | corroded | resist | resist | resist | resist | resist | resist | resist | resist | resist |
| Liquid Ga wettability | low | low | high | high | low | low | low | low | low | high |
| Acid test | 5 | 4 | 4 | 4 | 3 | 4 | 5 | 5 | 4 | 4 |

While the high RRR approach to the search of high quality Niobium coatings has not brought the expected success, another approach for searching high quality corrosion barriers has been pursued. So far as amorphous films have high corrosion-resistance due to the absence of typical structural micro-defects of the crystalline state (dislocations and grain boundaries) the small grain strategy was chosen, by exploring low temperature deposition, high rate sputtering, Nb alloying and Oxygen admixture.

Depositions have been carried out onto substrates cooled down to 22°C by water flux (*Nb 27*); cooled to 0÷10°C (*Nb 35, Nb 36*) by cold gas flux from pressurized liquid nitrogen; cooled to -50 ÷ -110°C (*Nb 34, Nb 32, Nb 33*) by liquid nitrogen flux. In the search-path toward amorphization, the low temperature deposition process has been found efficient for obtaining small-grain Niobium coatings. However, the grain size of Niobium coatings has a low limit of ~60 Å (Table 2). Even if the Niobium coatings obtained at low temperature have been found to resist to liquid gallium they have shown quite low resistivity to the acid solution (Figure 1, b). That should be explained by that fact that liquid gallium penetration depends both on the grains size, grain boundaries and maybe also roughness. Low temperature deposited Niobium coatings have been shown low wettability by liquid gallium, but it is not sufficient to prove their good barrier quality.

The increase of the deposition rate when cooling the substrate has not been found to be efficient for amorphization: high deposition rate, 5.5 nm/sec, $I_{DC}$=5.2 A (*Nb 25*) has no effect on grain size and on liquid gallium resistance, if compared with sample

Nb27, obtained in the same conditions (22 °C, $3 \cdot 10^{-2}$ mbar Ar pressure) but with lower deposition rate, 0.65 nm/sec, $I_{DC}$=0.5 A.

Moreover Niobium coating obtained at high deposition rate has been found more wettable by liquid gallium but more resistant to acid penetration. That could be maybe explained by the increasing of ionization degree of the particles impinging the substrate.

The increased sputtering gas pressure from $3 \cdot 10^{-3}$ mbar (*Nb 38*) to $3 \cdot 10^{-2}$ mbar (*Nb 39*) during Niobium thin film deposition caused the decreasing of the grain size from 170 Å to 110 Å (Table 1). However, the effect of gas pressure is less significant than the effect of cooling of a substrate.

### *Nb-Ti alloy deposition*

In order to test the influence of alloying on the microstructure of the coating and to prove that alloys result in better diffusion barriers than single metals, sputtering deposition of Niobium-titanium alloy using Nb-Ti alloy target (55% of Nb) at different argon pressure and substrate temperature has been carried out (Table 3).

Titanium is not a suitable element for the entrance foil coating, because of the number of isotopes produced under proton irradiation. However we adopted it just in order to investigate the influence of alloying on amorphization.

In order to avoid errors in alloy grain size evaluation, the shape of full normalized XRD spectrums has been used for interpretation of alloy films crystallinity (Figure 10).

In contrast to Niobium, the crystallinity of Niobium-titanium alloy films has been found dependent much more on sputtering gas pressure than on deposition temperature. Thus at $3 \cdot 10^{-2}$ mbar argon pressure amorphous-like coatings have been obtained at floating temperature (*Nb-Ti 4*) and by liquid nitrogen flux cooling at -50 °C (*Nb-Ti 7*). Similarly to pure Niobium high pressure sputtered films, Niobium-titanium coatings were found porous during the acid test. Nevertheless these alloy coatings have been found resistant to liquid gallium because of their low wettability that can be explained by high roughness. But then crystalline Nb-Ti alloys obtained at lower sputtering gas pressure (*Nb-Ti 2*, *Nb-Ti 5*) have been shown barrier properties extremely superior to all pure Niobium films obtained. Therefore Titanium cannot be used for entrance foils, but thanks to it we have understood that the presence in a coating of two types of atoms with different atomic radiuses could make the structure packing more compact.

**Table 3 Deposition of Nb-Ti alloy coatings**

| Sample Number | Nb-Ti 4 | Nb-Ti 7 | Nb-Ti 2 | Nb-Ti 5 |
|---|---|---|---|---|
| Temperature, °C | floating | -50 | floating | -50 |
| Ar pressure, mbar | $3 \cdot 10^{-2}$ | $3 \cdot 10^{-2}$ | $3 \cdot 10^{-3}$ | $3 \cdot 10^{-3}$ |
| $I_{DC}$, A | 0.5 | 0.5 | 0.5 | 0.5 |
| Deposition rate, nm/sec | 0.7 | 0.9 | 1.4 | 1.4 |
| Crystallinity due to XRD | amorphous-like | amorphous-like | crystalline | crystalline |
| Liquid Ga corrosion | resist | resist | resist | resist |
| Liquid Ga wettability | low | low | high | high |
| Acid test | 5 | 3 | 1 | 1 |

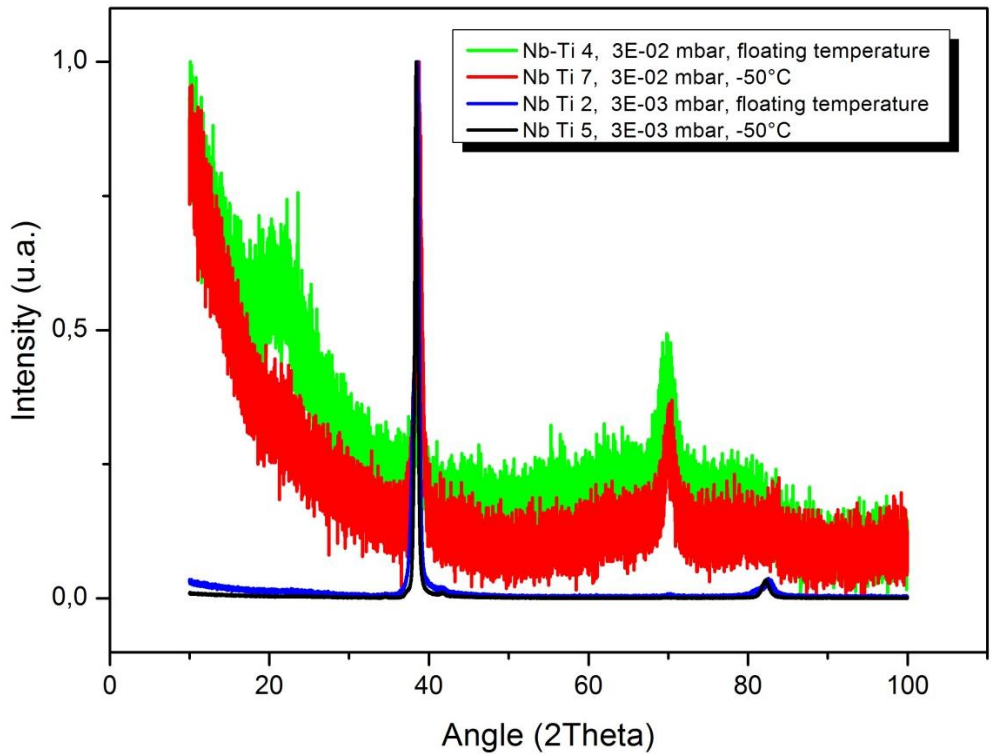

Figure 10 Influence of temperature and pressure on Nb-Ti alloy film deposition

### *Niobium Oxide*

On the basis of the literature approach for obtaining amorphous coatings, Oxygen admixture during Nb sputtering has been investigated.

While increasing the partial pressure of oxygen, the coating content changes from pure Niobium to Niobium with interstitial oxygen atoms and finally to pure $Nb_2O_{5-x}$. Further increasing of oxygen pressure doesn't bring changes into coating structure, but the excess of oxygen causes enhanced oxidation of Niobium target, which brings arcing and instability of the sputtering process. The stoichiometric conditions (for each working argon flux) have been found by varying the oxygen flux at constant argon flux and DC current. The right stoichiometry being found locking the oxygen flux at the point where the total pressure plot versus Oxygen flux changes slope and the voltage dependence on the oxygen flux has a jump. This method is rather qualitative, because it depends on sputtering system geometry, pumping system, substrate area and target poisoning. Nevertheless it was very useful for the fast achievement of almost stoichiometric Niobium oxide.

Table 4 Deposition of Niobium oxide coatings

| Sample Number | $Nb_2O_5$-1 | $Nb_2O_5$-2 | $Nb_2O_5$-3 | $Nb_2O_5$-4 | $Nb_2O_5$-8 | $Nb_2O_5$-6 |
|---|---|---|---|---|---|---|
| Temperature, °C | floating | floating | floating | floating | floating | floating |
| Bias, V | grounded | grounded | grounded | grounded | grounded | -100 |
| Sputtering gas pressure, | $6·10^{-2}$ | $7·10^{-2}$ | $3·10^{-3}$ | $3·10^{-3}$ | $1·10^{-2}$ | $1·10^{-2}$ |
| Ar flux, sccm | 25 | 25 | 1.7 | 1.7 | 3 | 3 |
| $O_2$ flux, sccm | 15 | 19 | 1.3 | 6 | 7 | 8 |
| $I_{DC}$, A | 0.5 | 0.5 | 0.5 | 0.5 | 0.5 | 0.5 |
| Deposition rate, nm/sec | 0.8 | 0.6 | 1.1 | 0.2 | 0.2 | 0.2 |
| Crystallinity due to XRD | amorphous | amorphous | crystalline | amorphous | amorphous | amorphous |
| Appearance | Metallic | Transparent | Metallic | transparent | transparent | transparent |
| Stoichiometry | Under | $Nb_2O_5$ | Under | $Nb_2O_5$ | $Nb_2O_5$ | $Nb_2O_5$ |
| Liquid Ga corrosion | resist | resist | resist | resist | resist | resist |
| Liquid Ga wettability | low | low | low | low | low | low |
| Acid test | 5 | 2 | 2 | 1 | 1 | 1 |

All stoichiometric Niobium oxide films have been found to be transparent and to have amorphous structure (Table 4). The crystallinity of under stoichiometric coatings which correspond to Niobium with interstitial oxygen atoms depends on the oxygen flux. As far the conditions will be from stoichiometry, as more crystalline the structure of the coating ($Nb_2O_5$-3) will be. The deposition rate of the stoichiometric oxide is lower than for the under stoichiometric coating. That can be explained by a higher degree of target oxidation. Negative bias was not found to have a substantial influence on Niobium oxide deposition.

All stoichiometric Niobium oxide coatings have been shown superior resistance both to the liquid gallium corrosion and to acid corrosion than pure Niobium films. The non-stoichiometric coatings have been found more porous, especially obtained at higher total gas pressure ($Nb_2O_5$-1).

All Niobium oxide coatings have been shown to resist to liquid gallium, show also low wettability.

Summarizing all results on Nb, Nb-Ti and $Nb_2O_5$ coatings (Table 5) stoichiometric niobium oxide obtained at lower sputtering gas pressure was found the most effective diffusion barrier.

Table 5 Summary of the experimental results obtained on Nb, Nb-Ti, $Nb_2O_5$, looking crystallinity/amorphization, resistance and wettability to Liquid Gallium and porosity to HCl, versus spattering conditions

| Methods: | | Nb | Nb-Ti | $Nb_2O_5$ |
|---|---|---|---|---|
| **Low Ar pressure** | XRD | crystalline | crystalline | Amorphous |
| | Liq. Ga test | resist, not wettable | resist, wettable | resist, not wettable |
| | Acid test | Porous | less porous | not porous |
| **High Ar Pressure** | XRD | crystalline | amorphous-like | amorphous |
| | Liq. Ga | Corroded | resist, not wettable | resist, not wettable |
| | Acid test | porous | porous | Porous |
| **Negative bias** | XRD | crystalline | crystalline | Amorphous |
| | Liq. Ga | resist, wettable | resist, wettable | resist, not wettable |
| | Acid test | less porous | less porous | not porous |
| **- 50 – 0 °C** | XRD | amorphous-like | depends on Ar pressure | Amorphous |
| | Liq. Ga | resist, not wettable | resist, not wettable | resist, not wettable |
| | Acid test | porous | porous | porous |

### Nb-Nb$_2$O$_5$ Multilayers

There are several reasons why the multilayer coatings could be better than the single layer of Niobium oxide:

- Niobium usually has better adhesion to a metallic substrate,
- thick layer of Niobium oxide has lower thermal conductivity than Niobium
- Niobium oxide is brittle.

So far as the grain size of Niobium depends on the thickness of a film, Niobium layers intercalated with Niobium oxide layers decrease the grain size of Niobium coatings. For multilayers, layer architecture is the most crucial parameter. Thus several multilayer samples with different periodicity were prepared, by fixing at 3 sccm the argon flux and periodically switching the oxygen flux from 0 sccm to 7 sccm.

The most significant samples are resumed in Table 6. We have investigated two totally different architectures of equal final thickness, preparing both coatings made by several thin layer (*M 1* - 30 double layers of 80 nm Nb plus Nb$_2$O$_5$) and coatings made by few thick layer (*M 10*, *M 11* - 5 double layers of about 400 nm Nb plus Nb$_2$O$_5$). In both cases, the Nb$_2$O$_5$ thickness is always ten times thinner than Nb.

Due to the big inertia of the pumping system when modulating the oxygen flux, broad interface between Niobium and Niobium oxide is expected.

**Table 6 Deposition of Nb-Nb$_2$O$_5$ multilayers**

| *Sample Number* | *M 1* | *M 8* | *M 10* | *M 11* |
|---|---|---|---|---|
| Temperature, °C | floating | floating | floating | floating |
| Gas flux control | manually | LabView | LabView | manually |
| Sputtering gas pressure, mbar | $3 \cdot 10^{-3}/1 \cdot 10^{-2}$ | $3 \cdot 10^{-3}/1 \cdot 10^{-2}$ | $3 \cdot 10^{-3}/1 \cdot 10^{-2}$ | $3 \cdot 10^{-3}/1 \cdot 10^{-2}$ |
| Ar flux, sccm | 3 | 3 | 3 | 3 |
| O$_2$ flux, sccm | 0/7 | 0/7 | 0/7 | 0/7 |
| I$_{DC}$, A | 0.5 | 0.5 | 0.5 | 0.5 |
| Average grain size, Å | 90 | 20 | 165 | 180 |

| Cubic lattice parameter, Å | 3.334 | 3.356 | 3.319 | 3.318 |
|---|---|---|---|---|
| Number of double layers | 30 | 60 | 5 | 5 |
| Total thickness, nm | 2300 | 5000 | 2300 | 2300 |
| Double-layer thickness, nm | 80 | 80 | 460 | 460 |
| Liquid Ga corrosion | resist | resist | corroded | resist |
| Liquid Ga wettability | low | low | low | low |
| Acid test | 2 | 1 | 2 | 2 |

The SEM picture (Figure 11), displaying the multilayer section of the 30 thin layer sample *M 1*, allows to estimate an average double layer thickness of about 60 nm.

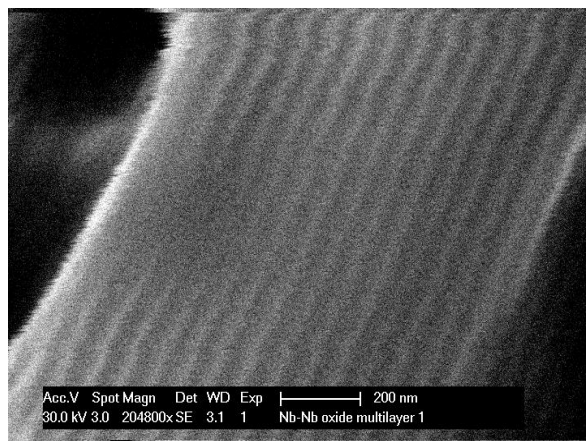

Figure 11 SEM picture of thin-layer Nb-$Nb_2O_5$ multilayer coating. Magnification 204800 X

In order to perform reliable multilayer depositions, a LabView program controlling the oxygen flux was used.

XRD investigation has clearly shown that the use of the automated program for controlling the gas resulted in finer grain microstructure compared with the manual control, especially in a case of thin-layered multilayer.

At constant total thickness, the decreasing of each double layer thickness has resulted in a significant decreasing of the average grain size of Niobium, passing from 165 Å for sample *M 10* to a 20 Å for sample *M 8*.

The multilayer Nb-$Nb_2O_5$ coatings have shown low wettability by liquid gallium,

thanks to the Niobium oxide top-layer. Also the resistance to the acid solution is similar to that of Niobium oxide and higher than the one of simple Niobium coatings.

While the thin-layered multilayer coating (*M 1*, *M 8*) show good resistance to liquid Ga penetration, the thick-layered multilayer coating (*M 10*) instead has been found to not resist. We suppose that so far as Niobium and Niobium oxide have had different thermal expansion coefficients a micro cracks had been able to appear during heating could allow liquid gallium penetration.

## Conclusions

Negative bias and heating applied to the substrate during deposition have been shown to increase the crystallinity of Niobium coating. Negative bias influence on Niobium microstructure for a magnetron sputtering has been found less effective than for a diode sputtering.

Low temperature deposition, increasing of sputtering gas pressure, admixture of oxygen into a metal film during deposition have been shown to be effective methods for decreasing of the grain size of sputtered Niobium coatings. Nevertheless these methods have not brought to pure amorphous films.

Among pure Niobium films, coatings obtained by applying -80 V bias to the substrate have been found to show the best resistivity to acid solution. However, all pure Niobium films have shown bad resistance to aqueous acid solution.

Crystalline Niobium-titanium alloy obtained at low sputtering gas pressure has been found resistant both for acid and liquid gallium penetration.

Sputtered amorphous Niobium oxide coatings have been found to be the best diffusion barrier, both for acid solution and for liquid gallium penetration.

Niobium-Niobium oxide multilayers were also found interesting for two reasons: the high thermal conductivity of Niobium is combined with the good barrier properties of Niobium Oxide; in addition the intercalation of Niobium Oxide does not allow the Niobium grain growth. In particular, thin layered multilayers have been found more resistant to temperature change.

**Acknowledgments**

This work was financed by the V Group of INFN for Accelerator and Interdisciplinary Physics, with a contribution from BEST Cyclotrons.

We are indebted with prof. G. Fiorentini, Dr. G. Cuttone and Prof. M. Carpinelli, Dr. Eng. L. Piazza for their precious support, and all the staff of the INFN-LNL Service for Material Science and Technology applied to Nuclear Physics for their constant help.